\documentclass[runningheads]{llncs}
\usepackage{graphicx}
\usepackage{amsmath}
\usepackage{amsfonts}
\usepackage{amssymb}
\usepackage{listings}
\usepackage{hyperref}
\usepackage{wrapfig}
\usepackage{array}
\usepackage[hyphenbreaks]{breakurl}

\begin{document}

\title{ Debugging Smart Contract's Business Logic 
		Using Symbolic Model Checking }
\titlerunning{Debugging Smart Contract's Business Logic Using Symbolic Model Checking}
\author{Evgeniy Shishkin}
\institute{InfoTeCS, Advanced Research Department \\
	Stariy Petrovsko-Razumovskiy Proezd 1/23 bld.1, Moscow, Russia \\ \email{evgeniy.shishkin@gmail.com}}

\maketitle

\begin{abstract}
Smart contracts are a special type of programs running inside a
blockchain. Immutable and transparent, they provide
means to implement fault-tolerant and censorship-resistant services. 
Unfortunately, its immutability causes a serious challenge of ensuring
that a business logic and implementation is correct upfront, before
publishing in a blockchain. Several big accidents have indeed shown 
that users of this technology need special tools to verify
smart contract correctness. Existing automated checkers are able to 
detect only well known implementation bugs, leaving the question of 
business logic correctness far aside.
In this work, we present a symbolic model-checking technique along with
a formal specification method for a subset of Solidity programming
language that is able to express both state properties and
trace properties; the latter constitutes a weak analogy of temporal
properties.
We evaluate the proposed technique on the MiniDAO smart contract, a young
brother of notorious TheDAO. Our Proof-of-Concept was able to 
detect a non-trivial error in the business logic of this smart contract
in a few seconds.


\end{abstract}
\section{Introduction}
	In 2008, Satoshi Nakamoto
	\footnote{This is a pseudoname. Real name of this 	crypto enthusiast is still unknown.} 
	published a paper where he described 
	an architecture of fully distributed decentralized paying system 
	called Bitcoin \cite{nakamoto08}.
	
	Bitcoin is a distributed ledger of user balances in essence. 
	What makes this system really unique is an inherited possession of 
	several properties, namely massive fault-tolerance, censorship-resistance, 
	authenticity of data and transparency of operations.

	A family of systems that inherits main Bitcoin traits is now known under
	a term blockchain. Shortly after Bitcoin gained traction, it became clear
	that the blockchain can be used not only as a distributed ledger of user
	balances, but also as a distributed computation platform that is able to
	execute some business logic by reacting on user inputs and maybe exchange
	value between participants in a form of crytocurrency transfers.

	The very first project that has implemented this idea was Ethereum
	 \cite{buterin13}.
	A business logic in its context is called a smart-contract, where user
	inputs are called transactions.

	Immutability and transparency of smart contracts provides never seen before
 	level of trust for end users enabling an emergence of new kind of business
 	models built on top of this technology. Unfortunately, its immutability
 	turns out
 	to be a threat in some cases: if an error sneaked in a smart contract, and
 	developers failed to find it before publishing the contract in a
 	blockchain then there will be no way to fix it afterwards, while an attacker
 	is able to exploit the error in any suitable moment.

	TheDAO attack \cite{dao} has demonstrated that this is indeed the case. 
	The attack led to money loss as big as 60 million dollars in crytocurrency.
    There were yet more several big accidents since then \cite{multisig}
    \cite{multisig2}
			
	This issue presents a serious threat to a wide technology adoption and 
	is currently well understood by the Ethereum community and other
	users.			
	At this moment, if you are a smart contract developer and would like
	to ensure that your contract is reliable, you have two options: either 
	apply for a manual human-driven audit in some blockchain security company 
	or use any of available automatic checking tools that scan your 	
	source code for typical error patterns such as integer overflows, 
	reentrancy bugs, transaction order dependency, etc. 
	The former option costs a lot of money,
	takes time and,  to the best of our knowledge, usually does not include
	business logic inspection, so errors in algorithms will stay.	
	The letter option is free, rather fast but unsound and incomplete: 
	in many cases you will get a lot of false positives, and a tool is 
	able to detect only publicly known implementation vulnerabilities,
	never inspecting business logic of a contract.

    In our opinion, for smart contract technology to gain mass adoption, 
    the community needs a tool that can automatically check that smart 
    contract implementation reflects the desired functionality, 
    and do it in a reliable cost- and time- effective way. 
    
    In this work, we present our effort towards building such tool. 
    We take Ethereum as a reference blockchain platform and Solidity as a 
    reference smart contract programming language. A formal specification of
    a contract can be given in several forms: global invariant on state
     variables, 
    properties on trace of a bounded length originating from initial state 
    and properties on chains of events. 
    
    We use SMT-solver as a back-end to find counter-examples of a given
    property. To make model-checking problem tractable, we consider only
    a subset of Solidity language without cycles, recursion and dynamic
    memory management.
    In order to evaluate the approach in practice, we conduct a 
    case study: we verify several functional properties of MiniDAO smart
    contract, a young brother of infamous TheDAO. In a few seconds our 
    Proof-of-Concept finds violation of some non-trivial temporal 
    property illuminating an error in a logic of the contract. 
    To the best of our knowledge, such inspection is out of rich for 
    existing automated tools.
 
	 \subsubsection{Main Contributions}   
	 \begin{itemize}
	 \item We formulate 4 classes of properties suitable for formal 
	 specification of smart contract behaviour, including temporal behaviour.
	 \item A subset of Solidity programming language called Sol is described. 
	 The subset is picked out with a purpose to make formal analysis easier.
	 \item A procedure of encoding Sol program and its formal specification 
	 into appropriate SMT formula is described.
	 \item A Proof-of-Concept demonstrating feasibility of proposed approach 
	 is implemented. Evaluation results are presented.
	 \end{itemize}

	Symbolic model-checking is a well-known method for verifying temporal
	properties of reactive systems on a bounded model. However, as far as we
	concerned, this work is among the first that applies this general method 
	to smart contracts verification. Moreover, we have not seen any 
	attempts to introduce classes of properties suitable for describing 
	functional specifications for this type of systems. Hopefully, 
	our work partially fills this gap. Results of implemented 
	Proof-of-Concept convince us that the approach is viable and deserves 
	a more thorough research.
	
\section{Ethereum Smart Contracts}

 	In this work we propose a method for verifying Ethereum smart
 	contract properties. We assume that contracts are programmed in a 
 	subset of  Solidity programming language. Due to
 	lack of space, we omit an introduction into the platform and the language.
 	We advise to read \cite{soliditydocs} for such introduction.
\subsubsection{Sol language.} 
	To make a model-checking procedure tractable in practice, we picked out 
	a subset of Solidity language that is rich enough
	to express interesting business logic without 
	producing astronomical number of intermediate states.

	Our small investigation into a typical program structure of a Solidity 
	smart contract shows that from 27341 publicly available smart contracts 
	\footnote{https://etherscan.io/contractsVerified, July, 2018}, only 23\% 
	uses
	any form of iteration and 26\% uses dynamic array structures. 
	However, it is well known that those
	programming constructs are among the most difficult ones when it comes to 
	model-checking, so we excluded them from our scope for now.
	
	Solidity language has a mapping data type which gives an ability to find an 
	element
	without iterating through a collection. This data type partially reduces 
	the dependency on cycles and recursion. It is interesting to note that one
	of the most cited smart-contracts, TheDAO \cite{jentzsch16}, uses no cycles or recursion in
	its implementation.
	
\subsubsection{Sol vs. Solidity.} 
	1) for, do/while, recursion constructs are forbidden 2) no dynamic memory
	 management support 3) only static arrays 4) var keyword is forbidden 5)
	 only 1 smart contract is allowed to be defined 6) events are allowed to
	 use no more than 4 arguments 7) address datatype	   
	 is given by a finite set of values 8) calling other contract's methods is
	 not allowed; dynamic smart contract creation is also forbidden. 9) bitwise
	 and string both datatype and operations are not supported at the moment
 
 \section{System Model}
 We assume that there is only one smart contract and a finite set of users
 performs transactions into that contract. This approximation is good enough
 for most practical use-cases.
 
 Let $N_{256} = \{0\,...\,2^{256}-1\}$, $Addr = \{a_0 \, ... \, a_n\}$. Lets
 suppose that every contract state variable has an associated unique natural
 number. We denote a set of all possible variable values as $Val$. Let $\Phi$
 denote a set
 of public functions of a contract, $E$ denote a set of events. Lets encode a
 system state as a triple: $\sigma = \langle \sigma_c, b, t \rangle $, where
  $\sigma_c$ is a contract state, $b:Addr \rightarrow N_{256}$ is a balances of
  blockchain addresses, $t:N_{256}$ is a time of last block related to the smart
  contract. We denote a set of all possible system states as $\Sigma$, so 
  $\sigma \in \Sigma$.
  We define smart-contract state as $\sigma_c = \langle \sigma_{cs}, alive,
  eventlog \rangle$, where $\sigma_{cs} :  \mathbb{N} \rightarrow Val$ denotes
  the current state of contract variables, $alive: \{True, False\}$ indicates if
  a contract is active or has been deleted, $eventlog:\{0\} \cup E$ denotes an
  event, generated by a contract during execution of last transaction, or
  absence of thereof. 
  
  Note that a function in Solidity is able to emit several
  events per transaction, and all of them will get into transaction log, but we
  intentionally
  limit our model to only one event per transaction, because, on one
  hand, it makes the model conceptually simpler, on the other, any finite chain
  of events can be encoded into some unique single event, so this restriction
  is not fundamental.
  
  Let us clarify the nature of variable $t$. Transactions in Ethereum 
  are executed
  by mining nodes not in FIFO order as one could imagine, but in blocks of 
  limited size. When block is full and ready to be processed, the node
  starts executing transactions within that block in some unspecified order.
  A timestamp is assigned to each new block upon its creation and is called
  $blocktime$. This value is monotonically increasing and is used a source 
  of relative time in business logic. It can be read from Solidity program 
  using $now()$ function call. So the value of $t$ refers to the value of 
  $blocktime$.
  
  Let us clarify a structure of set $\Phi$. In Solidity, a function
   definition consists of function name, function arguments, modifiers,
   return type and a function body. Modifiers can specify visibility 
   (external, public, etc) and also specify if a function is allowed to 
   accept cryptocurrency payments together with a function call.
  Taking it into consideration, for each function
  $$function\, f_i(arg_0, arg_1, ..., arg_n) \, public \, [payable]\, [return (T)]
  $$
  we construct a function $f_i'(\sigma_i, v, s, t, p)$, where $\sigma_i \in \Sigma$
  is a system state in the moment of a function call, $v \in N_{256}$ is an
  amount of cryptocurrency sent together with a call, $s \in Addr$ is a
  transaction sender address, $t \in N_{256}$ is a blocktime of a block that
  includes the transaction with that function call, $p = (arg_0,arg_1,...,arg_n)
  \in \Pi$ is a tuple consisting of function formal arguments, so $\Phi =
  \bigcup\limits_{i=1}^{n} \{f'_i\}$. The function body of $f'$ is generated
  from body of $f$ by making a series of substitutions: we substitute $now()$
  for $t$, $msg.sender$ for $s$, $msg.value$ for $v$, and so on. We also change 
  the return value of a function: instead of returning a value of type T,
  a function $f'_i$ returns a new system state $\sigma_{i+1}$. The returning
  value is omitted for now, as it is rearly used by external agents in practice,
  because it is rather difficult to monitor it. A mechanism of events is usually
  used instead.
  
  We shall define several notions that let us model a system executing a 
  smart-contract in time.
  
  \begin{definition} (Initial states). A set of initial states is defined as follows:
 
    $ I = \{ \{ \sigma_c^0, b_0, t_0 \} \, | \, b_0:Addr \rightarrow N_{256},
    t_0 \in N_{256} \}, \sigma_c^0 = \{ \sigma_{cs}^0, True, 0\} $, where $b_0$
    is  a map between addresses and balances, $t_0$ denotes the blocktime of a
    block
    where the contract has been created, $\sigma_{cs}^0$ is a contract state
    right after a successful constructor call.
    \end{definition}
   
   In Sol, it is prohibited to use any expressions that can lead to exceptions 
   inside a constructor, along with $transfer$ and $selfdestruct$ function calls.
   That is why we assume that a constructor call is always successful, and its
   side effect is expressed solely as assignments of values to variables of the
   contract.
   
   \begin{definition} (Trace). Any finite sequence of system states 
   $\sigma = \sigma_0 \sigma_1 ... \sigma_{k-1}$ , 
   $\sigma_i \in \Sigma$ is called a trace iff the following holds: 
   $$trace(\sigma) = \forall i \in \mathbb{N}, 
   0 \leq i < len(\sigma) . \delta(\sigma_i, \sigma_{i+1})$$
   where $\delta(\sigma_i, \sigma_{i+1})$ is called a step relation (defined
   below), $\sigma_0 \in I$, $len(\sigma)$ denotes a sequence length.
   
   \end{definition}
	
	We denote a set of all possible state sequences (not only traces) 
	as $2^\Sigma$.
	
	Generally, a smart contract can transition into several different states
	from its current state because it is not known in advance what function
	with what parameter values and from what user address will be called. 
	Smart contract is a highly non deterministic system reactive system. That is
	why when we reason about a contract we need to reason about all possible
	traces of that contract, not to loose interesting states that can contain
	an error.
	
	\begin{definition} (Behaviour). Let $\Sigma^{*} \triangleq
	\{ \sigma \, | \, \sigma \in 2^\Sigma \wedge trace(\sigma) \}$
	is a set of all system traces. We call $\Sigma^{*}$ a behaviour of a
	system.	
	\end{definition}
	
	We are allowed to put $\sigma_{i+1}$ after $\sigma_i$ only if a pair
	$( \sigma_i, \sigma_{i+1})$ is in a special relation that we 
	call a step relation. To define this relation, we need to clarify
	what exactly smart contract transaction does.
	
	\subsubsection{Partial functions.} Generally speaking, a function $f' \in \Phi$ is not total: its execution can lead to a phenomena called an exception.
	Exception is an interruption of a function execution due to executing some
	forbidden or undefined operations. Exceptions result in a complete
	side-effect rollback, including cryptocurrency transfers, changing variable
	values, etc. Functions that can lead to exceptions are called partial, 
	because they are defined only on a subset of all possible values of its
	arguments and external parameters. We could completely ignore this
	phenomena by making the $\delta(\sigma_i,\sigma_{i+1})$ relation a reflexive
	one. But, in this case, we would get a huge amount of garbage transitions
	that do not lead to new states. This would greatly reduce an efficiency of 
	symbolic model-checking procedure. To exclude such transitions we 
	introduce a notion of function precondition.

	\begin{definition} (Function precondition). A precondition of a function\\
	$f'(\sigma_i,v,s,t,p) \in \Phi$ is a predicate 
	$f_{pre}(\sigma, v, s, t, p)$ such
	that if the predicate holds for some $\sigma \in \Sigma, v \in N_{256},
	s \in Addr, t \in N_{256}, p \in \Pi$ then $f' \in \Phi$ is defined on
	those arguments. If we equip every public function with such predicate, 
	we get $\Phi_{pre} = \{ (f', f_{pre})\}$
	\end{definition}

	\begin{definition} (Step relation). It is possible to make a step from 
	a state $\sigma_i$ into a state $\sigma_{i+1}$ if there exist a set of 
	arguments $v, s, t, p$ such that at least one function $f' \in \Phi$ 
	is defined on those arguments $\sigma_i, v, s, t, p$.
	A set of all such pairs of states we call a step relation:
	$$\Delta = \{ (\sigma_i, \sigma_j): \exists (f',f_{pre}) \in \Phi_{pre},
	v, s, t, p \, . \, \sigma_j = f'(\sigma_i, v, s, t, p) \wedge f_{pre}(\sigma_i, v, s,
	t, p) \}$$	
	\end{definition}
	For convenience, we define
	$\delta(\sigma_i, \sigma_j) = ( (\sigma_i, \sigma_j) \in \Delta)$.
	
\section{Verification problem}
To this point, we have defined a model of interaction between user and a smart-contract. Lets define a problem we are aiming to solve.

\begin{definition} (Verification problem). 
Assume $P : \Sigma^{*} \rightarrow \mathbb{B}$ is a predicate over a
 trace. We have to show, that 
 $\forall\,  \sigma \in \Sigma^{*} \, . \, \sigma \models P$ that is
 to show that a behaviour of a contract obeys $P$. We call $P$ a formal
 specification in this context.
\end{definition}

Depending on a type of functional property, predicate $P$ can take a single
state (trace of length 0) $P:\Sigma \rightarrow \mathbb{B}$ or a trace of fixed length 
$P:\Sigma_k^{*} \rightarrow \mathbb{B}$ as its argument. We define
$\Sigma_k^{*} = \{ \sigma \in \Sigma^{*}\, | \, len(\sigma) \leq k \}$, where
$len(\sigma)$ is a length of a state sequence.

Lets define several types of functional properties we would like to be able
to check. 

\begin{definition} (Invariant). A predicate $P:\Sigma \rightarrow \mathbb{B}$
is called an invariant iff 
$$ \forall \sigma_0 \in I\, . \, P (\sigma_0) \, \wedge \, \forall \sigma_i,
 \sigma_j \in \Sigma \, . \, (\delta(\sigma_i, \sigma_j) \wedge P (\sigma_i))
 \rightarrow P (\sigma_j) $$
\end{definition}
\begin{definition} (Trace property of length k). A predicate 
$P:\Sigma \rightarrow \mathbb{B}$ is called a trace property of length k iff
$$ \forall \, \sigma^{*} \in \Sigma_k^{*}, i \in \mathbb{N} \, . \, i \leq len(\sigma^*) \rightarrow P (\sigma_i^*) $$
\end{definition}

We introduce another class of properties that we call \textit{events chaining
 properties}.
Properties of this type are given by a predicate over a trace of length $k$,
so $P: \Sigma_k^* \rightarrow \mathbb{B}$. We shall define one instance of
this property class, others can be defined similarly. We could not generalize
our definition to a greater extent at this point.

\begin{definition} (Events chaining)
If, during an execution, an event \\ $E_1$ has occurred and $E_2$ has occurred afterwards, then there must be no event $E_3$ emitted in between, i.e.\\
$$P(\sigma^*) = \exists \, i,j \in \mathbb{N} \, . \, \sigma_i^*[eventlog] = E_1
 \wedge \sigma_j^*[eventlog] = E_2 \wedge i < j 
 \leq len(\sigma^*) \rightarrow $$ 
$$\forall k \in \mathbb{N} . i < k < j \, . \, \sigma^*_k[eventlog] \neq E_3 $$
\end{definition}
Extra dependencies between event arguments $p_0,...,p_i;m_0,...,m_j;n_0,...,n_k$
and arguments of a function being called $\sigma, v, s, t, p$ must be given
as an extra predicate conjuncted with the one defined above.

	The following class of properties express the idea that between events $E_1$
and $E_2$ is it always possible (or impossible) to perform a function call into the smart
contract with a given arguments. We call this class of properties as
\textit{function call possibility}. As in the previous case, only one instance
of such property class is given.

\begin{definition} (Function call possibility)
Let $P: \Sigma_k^* \rightarrow \mathbb{B}$. If an event $E_1$ has occured and
$E_2$ has occurred afterwards, then it must be the case that a function call
$f'(\sigma, v, s, t, p)$ must be successful at any state between them.
$$P(\sigma^*) = \exists \, i,j \in \mathbb{N} \, . \, \sigma_i^*[eventlog] = E_1
 \wedge \sigma_j^*[eventlog] = E_2 \wedge i < j 
 \leq len(\sigma^*) \rightarrow $$
$$ \forall k \in \mathbb{N} , i < k < j \, . \, f_{pre}(\sigma_k^*, v, s, t, p)
 = True $$
\end{definition}

\section{Model Construction}
To construct a model of a contract means to be build such tuple:\\
$ (\Phi_{pre}, E, Addr, I, k, \Sigma_k^*, P) $, where $k$ denotes a 
path length. Lets briefly discuss how each of those
elements are being built for a given smart-contract.

\paragraph{\underline{Set Addr}.} Ethereum has $2^{160}$ distinct addresses available to 
be used, but for symbolic model-checking such set would be unmanageable: its
size influences a number of ways a transition can be made from one state into
another. 
That's why we try to choose optimal set $\{a_0, ..., a_n\}$ such that it enables
most principal scenarios of smart contract to be modeled. This choice is not
automated in any way at this point, by default it is set to 
$Addr=\{noAddr, addr_0, addr_1, addr_2, contractAddr \}$.
The element $noAddr$ denotes 'undefined' address value, a thing that is 
usually coded as $address(0)$ in Solidity. The element $contractAddr$ denotes
an address of the contract being executed.
\paragraph{\underline{Set $\Phi_{pre}$}.} This set is being built automatically:
for every
public function $f'$ of smart contract (except a constructor), we build 
$f_{pre}(\sigma, v, s, t,  p)$. It is done by symbolically executing a function
$f'$ and building a set of constraints that do not allow $f'$ to throw an
exception. All function calls to other internal functions are preprocessed by
inserting a function body into the function $f'$.

A list of expressions that are able to throw an exception is: division operator;
reminder; mulmod; addmod; transer; assert, require, revert functions; throw;
calling non-payable functions with $v > 0$; a situation when $s = contractAddr$;
trying to execute a function of a deleted smart contract, i.e. 
$\sigma_{cs}[alive] = False$. 

There are no cycles and recursion available in Sol language, that's why we are
guaranteed for this symbolic execution procedure to terminate.

\paragraph{\underline{Set $E$}.} This set is being built automatically from
events defined
in the smart contract. Events that are not used will be omitted.
\paragraph{\underline{Set $I$}.} A set of initial states. The set is being built
automatically by evaluating a constructor of the smart contract.
\paragraph{\underline{Value $k$}.} The value denotes an exact length of a trace.
It is given explicitly by the user. If a property being checked is an
invariant,then this parameter is ignored.
\paragraph{\underline{Set $\Sigma_k^*$}.} The set is constructed implicitly, by
giving a system of contraints over a set of system states $\sigma_{[0...k-1]}$
and a set of parameters $(v, s, t, p)$, where i-th tuple of parameters
corresponds to the i-th function call. Lets define a transition relation as
follows: 
\begin{align*}
trans&ition(\sigma_i, \sigma_{i+1}) = \\
 &\bigvee\limits_{j=1}^{| \Phi_{pre} |}\
{(f^{pre}_j(\sigma_i, v_i, s_i, t_i, p_i) = True \wedge \sigma_{i+1} =
 f'_j(\sigma_i, v_i, s_i, t_i, p_i))}
\end{align*}

\begin{definition}(Path). Lets define a path between states as follows:
$$path(\sigma_{[0..k]}) = \bigwedge\limits_{0 \leq i < k}{transition(\sigma_i
,\sigma_{i+1})}$$
\end{definition}

The difference between a path and a trace is in the originating state: 
in the former, it is allowed to be any state, not only initial. A path of length
zero consists of only one state, and has no transitions.

Let us introduce a time monotonicity requirement: 
$T = \bigwedge\limits_{i=0}^{k-1}{t_i < t_{i+1}}$, and a restriction on
 originating addresses: 
$$NoSelfAddrCall = \bigwedge\limits_{i=0}^{k-1}{(s_i \neq contractAddr \wedge
s_i \neq noAddr)},$$ and a requirement for initial state origin: $I(\sigma_0)$.
In
 this case, a predicate \\ $I(\sigma_0) \wedge T \wedge NoSelfAddrCall \wedge 
 path(\sigma_{[0..k-1]})$ describes a transition system of our smart contract
 executing all possible $k$ consecutive transactions.
Finding the right assignments for variables $\sigma_{[0..k-1]}$ and $(v,s,t,p)$ of this formula inside an SMT-solver implicitly generates the set $\Sigma_k^*$.

\paragraph{\underline{Predicate $P$}}. The predicate is given by the user and
describes formal specification for the contract.

Thereby, it is possible to automatically extract a model from a Sol program 
that can be checked using an SMT-solver. A user has to specify a 
functional specification $P$, trace length $k$ and the number of 
distinct addresses in the set $Addr$.

\section{Verification algorithms} \label{algorithms}
Here, we describe algorithms that check aforementioned classes of properties.
Let $P(\sigma)$ denote a property we would like to check. 
The expression $SAT(e,Vars)$ means that $e$ is checked by SAT/SMT solver for
satisfiability, i.e. a solver look for values for variables of $Vars$ such that
expression $e$ evaluates to $True$. If such assignment is found then the
expression returns $True$, otherwise $False$. A result $unknown$ is not 
concerned here because we stay inside decidable theories, and for such theories
this answer simply means that a solver ran out of time. 
A predicate $path(\sigma_{[0..n]})$ is defined the same as above. 
Each algorithm returns $True$ if property $P(\sigma)$ holds, otherwise 
$False$ is returned. 
   
In the pseudo-code, the expression $Vars = \{ p_i:t_i \}$ means that for
every variable $p$ we add corresponding proposition variable of type $t$
or an array of variables into solver's context.

Some of those algorithms were published in the past. The model-checking 
procedure for paths of length $k$ was thoroughly described in different
variations in \cite{sheeran2000}. The invariant proving algorithm is 
considered to be well known also. However, to make the work self-sufficient, we
include pseudo-code for those algorithms also.

\begin{lstlisting}[frame=single,
				   caption=Algorithm 1. Checking invariant property,
				   numbers=left,
				   escapeinside={(*}{*)}]
Vars = {(*$\sigma_{[0,1]}:\Sigma$, 
$v_{[0,1]}:N_{256}$, $s_{[0,1]}:Addr$, $t_{[0,1]}:N_{256}$, $p_{[0,1]}:\Pi$*)}
if (SAT((*$I(\sigma_0) \, \wedge \, \lnot P (\sigma_0),\,Vars) $*)) {
   print (*$s_0$*)
   return false
}
if (SAT((*$P(\sigma_0) \wedge \delta(\sigma_0,\sigma_1) \wedge \lnot P(\sigma_1)), Vars))$*) {
   print (*$f', \sigma_0, \sigma_1$*)
   return false
}
return true
\end{lstlisting}

\begin{lstlisting}[frame=single,
				   caption=Algorithm 2. Checking a trace property of length k,
				   numbers=left,
				   escapeinside={(*}{*)}]
Vars = {(*$\sigma_{[0..k-1]}:\Sigma$,\,
$v_{[0..k-1]}:N_{256}$, $s_{[0..k-1]}:Addr$,\,
$t_{[0..k-1]}:N_{256}$*),
        (*$p_{[0..k-1]}:\Pi$*)}
i = 0
while (i < k) do {
  if (SAT((*$I(\sigma_0) \, \wedge \, path(\sigma_{[0..i]}) \wedge \lnot P(\sigma_i) ,\,Vars) $*)) {
     print (*$\sigma_{[0..i]}$*)
     return false
  }
  i = i + 1
}
return true
\end{lstlisting}

\begin{lstlisting}[frame=single,
				   caption=Algorithm 3. Checking events chaining property,
				   numbers=left,
				   escapeinside={(*}{*)}]
Vars = {(*$\sigma_{[0..k-1]}:\Sigma$,\,
$v_{[0..k-1]}:N_{256}$, $s_{[0..k-1]}:Addr$,\,
$t_{[0..k-1]}:N_{256}$*),
        (*$p_{[0..k-1]}:\Pi, m, n, q: N_{256}$*)}
i = 3
while (i < k) do {
  if (SAT((*$I(\sigma_0) \, \wedge \, path(\sigma_{[0..i]}) \wedge 
      \sigma_m^{cs}[eventlog] = E_1 \wedge
      \sigma_n^{cs}[eventlog] = E_2 \, $*)
          (*$\wedge \, (m < n) \wedge (q > m) \wedge (q < n) \wedge
          \sigma_q^{cs}[eventlog] = E_3,
  \,Vars) $*)) {
     print (*$\sigma_{[0..i]}$*)
     return false
  }
  i = i + 1
}
return true
\end{lstlisting}

\begin{lstlisting}[frame=single,
				   caption=Algorithm 4. Checking functional call possibility property,
				   numbers=left,
				   escapeinside={(*}{*)}]
Vars = {(*$\sigma_{[0..k-1]}:\Sigma$,\,
$v_{[0..k-1]}:N_{256}$, $s_{[0..k-1]}:Addr$,\,
$t_{[0..k-1]}:N_{256}$*),
        (*$p_{[0..k-1]}:\Pi, m, n, q: N_{256}$*)}
i = 3
while (i < k) do {
  if (SAT((*$I(\sigma_0) \, \wedge \, path(\sigma_{[0..i]}) \wedge 
      \sigma_m^{cs}[eventlog] = E_1 \wedge
      \sigma_n^{cs}[eventlog] = E_2 \, $*)
          (*$\wedge \, (m < n) \wedge (q > m) \wedge (q < n) \wedge
          \lnot f_{pre}(\sigma_q, v_q, s_q, t_q, p_q),
  \,Vars) $*)) {
     print (*$\sigma_{[0..i]}, v_q, s_q, t_q, p_q$*)
     return false
  }
  i = i + 1
}
return true
\end{lstlisting}

\section{Proof Of Concept}

In order to evaluate described algorithms, we have programmed the
MiniDAO smart-contract, a simpler version of infamous TheDAO, and tried
to find a counter-example of several specified requirements. 
The MiniDAO is programmed
using Sol programming language, and functional properties are encoded
in a first-order logic predicate form. Both artefacts (source code and
requirements) were manually translated into the language of SMT-solver and
model-checked, making time measurements for different properties using
different model parameters. We briefly describe the purpose of this contract
and some desired functional properties that we later use as a partial 
functional specification.

\subsection {MiniDAO Smart Contract}
MiniDAO is a smart contract that implements an autonomous investment fund.
There are two types of users in the fund: investors and contractors. Investor is 
a user that deposits his money into the fund and later votes for or against a
contractor's proposal. Contractor is a user that makes a proposal of some
new project and, if enough funds are raised, implements this project.
It is assumed that the invested sum and dividends are paid back to investors
later using smart contract facility, but this functionality is not 
implemented right now.
 
An interface of the MiniDAO is given on . The full listing is available at
 \footnote{\url{https://bitbucket.org/unboxed_type/minidao/src/master/contracts/MiniDAO.sol}}
 
To give investor an ability to get back his invested funds, the \textit{refund} method
is implemented. Refund is made only if the investor has not voted for any
proposal. Investors are able to transfer their tokens to other investors using
now standard ERC20 token interface functions.
\lstset{basicstyle=\footnotesize\ttfamily,breaklines=true}

\begin{lstlisting}[frame=single,
				   caption=The MiniDAO interface]
interface ERC20Interface { /* standard definition */ }				   
interface MiniDAOInterface {
  function deposit() public payable;
  function vote(uint proposalId, bool supports) public;
  function refund() public;
  function propose(address recipient, uint amount, 
  	string test);
  function execute_proposal() public;
  event Voted(address voter, uint proposalId, bool supports);
  event Refund(address investor, uint tokens);
  event Deposited(address investor, uint tokens);
  event ProposalAdded(uint amount, uint proposalId);
  event ProposalExecuted(uint proposalId);
  event ProposalRejected(uint proposalId);
}
contract MiniDAO is MiniDAOInterface, ERC20Interface { 
  /* Implementation goes here */ }
\end{lstlisting}

Suppose that we are interested in engaging as many investors as possible in
our MiniDAO fund. To minimize investors fear of money loss, we claim
the following: \textit{"If you do not vote for any proposal, you will 
always be able to refund your deposit"}. To prove our assertion, we show
the code of the \textit{refund} function.
\setlength\intextsep{0pt}
\begin{wrapfigure}[11]{l}{0.72\textwidth}
\vspace{-2pt}
\centering

\begin{lstlisting}[frame=single,
				   caption=The refund function body]
function refund() public {
  address sender = msg.sender;
  uint tokens = balance[sender];
  require (isVoted[0][sender] == false);
  require (tokens > 0);
  require (DAO_tokens_emitted >= tokens);
  DAO_tokens_emitted -= tokens;
  balance[sender] = 0;
  sender.transfer(tokens * DAO_token_price);
  emit Refund(sender, tokens);
}				   
\end{lstlisting}
\end{wrapfigure}
The function looks simple and convincing. Yet, our assertion does not hold.

\subsubsection{Majority Attack} Lets discuss a possible attack scenario. 
Suppose that two investors have invested crypto currency equal to X and Y 
MiniDAO tokens respectively. After that, the third investor comes in and invests
amount equal to $2 \times (X + Y)$ tokens. That third investor now has a majority
of votes ($2 / 3 \approx 66\%$) when decision is to be made about project
funding. It is not prohibited for an investor to be a contractor also, this
gives an opportunity for that investor to register his own proposal with
expected amount of as much as $3 \times (X + Y) $ tokens. After that, because
his vote is final, he votes for his own proposal and calls
 \textit{execute\_proposal}. All funds of MiniDAO are transferred to that
 third intruder investor leaving two other investors with nothing. We have
 an obvious violation of the aforementioned property: investors did not vote, but
 refund is now impossible for them.
 
 The Majority Attack is described in the TheDAO whitepaper \cite{jentzsch16}. 
 \textit{Lets suppose we do not know about this attack and would like
 to ask our verification tool to check the refund property 
 automatically}.

\subsection {MiniDAO Functional Requirements}
We give a partial functional specification by formulating 3 properties that we
will later check.
\subsubsection{NotVotedRefund Property.} If some investor with the
 address $inv$ deposited some amount of money and not voted for any proposal
 afterwards, then he will always be able to get a refund by calling the
 \textit{refund} function.
\begin{gather*}
NotVotedRefund(\sigma) = \exists i, inv, s, 1 \leq i < k . 
\sigma_{cs}^i[logs] = Deposited(inv, s) \, \wedge \\
\exists j, inv_1, s_1, id, 
1 \leq j < k . \sigma_{cs}^j[logs] = ProposalAdded(inv_1, s_1, id)\, \wedge \\
\forall n, id_1, 1 \leq n < k . 
\sigma_{cs}^n[logs] \neq Voted(inv, id_1, True) \, \wedge \\
\sigma_{cs}^n[logs] \neq Voted(inv, id_1, False)\, \wedge \\
\sigma_{cs}^n[logs] \neq Refund(inv) \wedge \sigma_{cs}^n[logs] \neq 
Transfer(inv) \rightarrow \\
\forall m, i \leq m < k \, . \, refund_{pre}(\sigma_m, v, inv, t, p)
\end{gather*}
\subsubsection{InvDaoBalance Property.} In any reachable state, 
the sum of MiniDAO token balances is equal to the total amount of emitted
 tokens, i.e.
$$ InvDaoBalance(\sigma_k) = \sum\limits_{i \in Addr}{\sigma_{cs}^k[daoBalance[i]] = daoTokensEmitted} $$
\subsubsection{RejectedNotExecuted Property}. A rejected proposal 
must not receive any funds from the MiniDAO.
\begin{gather*}
RejectedNotExecuted(\sigma) = \forall n . \exists i,j \in N .
\sigma_{cs}^j[logs] = ProposalAdded(n, amount) \, \wedge \, \\
\sigma_{cs}^j[logs] = ProposalRejected(n) \, \wedge
\, i < j \rightarrow
\forall k, i < k < j. \\ 
\sigma_{cs}^k[logs] \neq ProposalExecuted(n) 
\end{gather*}

\subsection{Finding Errors in MiniDAO}
As we discussed earlier, to construct a model means to construct
a number of objects $( \Phi_{pre}, E, Addr, I, k, \Sigma_k^*, P)$.
After that we can run one of verification algorithms given in section
\ref{algorithms}. The tool that we are building will construct those
objects automatically except $Addr$, $k$ and $P$. While it is not ready,
we constructed those objects manually. We translated MiniDAO source code
together with its specifications into the language of SMT-solver.

After this translation and some further optimizations, we received a model 
that produced a counter-example for two out of three properties during
model-checking.

\setlength\intextsep{0pt}
\begin{wrapfigure}[10]{r}{0.55\textwidth}
\vspace{-8pt}
\centering
\begin{lstlisting}[frame=single]
0. NoEvent
1. Deposited(addr4, 2976)
2. Deposited(addr2, 1672)
3. ProposalAdded(addr4, 4648, 1)
4. Voted(addr4, 1, 1)
5. ProposalExecuted(1)

step = 5, investor = addr2
\end{lstlisting}
\caption{The counter-example for NotVotedRefund property.}
\label{example}
\end{wrapfigure}

\paragraph{NotVotedRefund property.} The counter-example
\ref{example} is represented as a chain of occurred events. The first NoEvent event denotes
the initial state. The investor with address $addr4$ is not able to refund
after event 5. We did not fix this error in any way.

\paragraph{InvDaoBalance property.} This property was also violated. In the
first version of MiniDAO, the method $vote$ was wrongly settings the balance 
of investor to zero after the vote. The error was fixed and checked again.
After the fix, no counter-example was found.
\paragraph{RejectedNotExecuted property.} The check was successful, there was
no trace found that violated this property.

\section{Efficiency Evaluation}
We tried our proof-of-concept for different set sizes and path length,
measuring the time needed to find a counter-example or prove the absence
of thereof; results are presented in the table \ref{results} below.

Keep in mind that those results give only very rough presentation of
the model-checking efficiency for such systems because SAT/SMT-solvers
are known to be very sensitive to any changes in the model or in
the solver's parameters.

Experiments were conducted using Intel Core i7-4770 4 cores 3.4GHz CPU,
32 GB RAM running Linux Ubuntu 18.04.1 LTS 64-bit OS virtual machine 
using VirtualBox 5.1.24 hypervisor on Windows 7 OS. 
We have used Z3 SMT-solver ver. 4.8.2 64 bit, the model is written using 
Z3Py binding for Python. 
\\
\begin{figure}
\caption{Left table: a duration of checking NotVotedRefund property.
 Notation $>N$ means that we have stopped the checking process after N
  seconds. Middle table: a duration of checking RejectedNotExecuted
   property. Right table: a duration of checking InvDaoBalance property.
 Trace length column denotes the minimal length of a trace; the
 maximum trace length was set to 12 in all cases.
    }
\label{results}    
\begin{center}
\begin{tabular}{|>{\centering\arraybackslash}p{1cm}
				|>{\centering\arraybackslash}p{1.2cm}
				|>{\centering\arraybackslash}p{1cm}|}
\hline
Trace length&  
Integer width, bits & 
Time, sec\\
\hline
6 & 16 & 57 \\
\hline
8 & 16 & 10 \\
\hline
12 & 16 & 971 \\
\hline
6 & 32 & 47 \\
\hline
8 & 32 & 171 \\
\hline
12 & 32 & $>$ 2500 \\
\hline
\end{tabular}
\quad
\begin{tabular}{|>{\centering\arraybackslash}p{1cm}
				|>{\centering\arraybackslash}p{1.2cm}
				|>{\centering\arraybackslash}p{1cm}|}
\hline
Trace length &  
Integer width, bits & 
Time, sec\\
\hline
6 & 16 & 14 \\
\hline
8 & 16 & 11 \\
\hline
12 & 16 & 1.6 \\
\hline
6 & 32 & 48 \\
\hline
8 & 32 & 35 \\
\hline
12 & 32 & 5 \\
\hline
\end{tabular}
\quad
\begin{tabular}{|>{\centering\arraybackslash}p{1.5cm}
				|>{\centering\arraybackslash}p{1cm}|}
\hline
Integer width, bits & Time, sec \\ 
\hline 
16 & 637 \\
\hline
24 & 478 \\
\hline
- & - \\
\hline	
- & - \\
\hline	
- & - \\
\hline	
- & - \\
\hline	

\end{tabular}
\end{center}
\end{figure}

\section{Related Works}
A lot of research is being done in the area of smart contracts
 reliability, just to name a few \cite{atzei17}\cite{bhargavan16}
\cite{delmolino16}\cite{pettersson15}
\cite{luu16}\cite{kalra18}
\cite{hirai17}\cite{mueller18}.

In \cite{atzei17}, authors present a thorough list of known 
Solidity language and Ethereum Virtual Machine vulnerabilities.
TheDAO attack scenario is described in details. 

In \cite{bhargavan16}, authors propose a way to prove the absence of
several kinds of typical implementation errors by encoding Solidity* 
program info F* language together with some specially crafted abstract
data types that help track error codes management. 

In \cite{pettersson15}, authors investigate ways to program Ethereum 
smart contracts using Idris functional language; by using its powerful
type system, authors propose several algebraic types that help eliminate
typical implementation errors during a compilation stage. The back-end that 
translates Idris into EVM bytecode is also presented.

Works \cite{luu16} \cite{mueller18} describe tools for Solidity smart contract
static analysis called Oyente and Mythrill accordingly. Tools are based on
symbolic execution technique and aim to find a series of conditions that, when
met, will result in executing potentially harmful programming constructs. 

In \cite{kalra18}, a tool called ZEUS is presented. This tool can find 
a scenarios of deviation from user-specified policies. Policies are written
in a language of assertions about smart-contract state variables with 
basic arithmetic support. In this respect, ZEUS is the closest tool to ours
in that it makes an effort to analyse business logic against specification
 stated in a form of policies.
Unfortunately, ZEUS is not able to analyse temporal properties, and is aiming to
check only state properties.

The work \cite{hirai17} is among the first that proposed using a theorem prover
environment to formally verify properties of a smart-contract. The author 
encoded EVM instructions semantics in Coq and later in Isabelle. 
Using this framework, a developer is able to prove properties of his 
smart-contracts on the bytecode level inside a theorem prover. 
Being the most foundational among all other approaches, this method gives the
highest guarantees on correctness of producing artefacts, but potentially takes
a lot of time and effort.

One more step in similar direction is \cite{zakrzewski18}. In this work, 
an effort towards formalizing a subset of Solidity language semantics 
inside a theorem prover is described. 

\section{Conclusion and Future Work}
In this work, we described a conceptual device of symbolic model-checker
for Solidity smart-contracts based on SMT-solver, that is able to detect
deviations from prescribed
behaviour on traces of the given length, or breaking an invariant. 

While the model-checking tool itself is under construction, the viability of the
proposed method was evaluated on the MiniDAO smart-contract, a weak analogy of
TheDAO. Results convinces us that proposed method is indeed practical and 
needs further investigation.
For the future work, we would like to mention the following directions:
\begin{itemize}
\item The specification language need not be as complicated as 
first order logic language, which is quite verbose and a bit alien to many
developers. We need to develop an alternative way of encoding behaviour 
of a smart-contract. Different flavours of temporal logic do not answer
this question in our opinion.
\item SMT-solvers are very sensitive to a model representation. Optimality
of such representation influences on how fast the model-checking process
will execute. Thus, we need to develop optimization strategies for representing
Solidity datatypes and code when expressed in SMT solver language.
\end{itemize}

\bibliographystyle{splncs04}
\bibliography{solsym}
\end{document}